# Radiation of a particle performing helical motion in a multilayer cylindrical waveguide

M.I. Ivanyan[1], B. Grigoryan[1], A. Grigoryan[1,2], L. Aslyan[1], V. Avagyan[1,3], H. Babujyan[1], S. Arutunian[1,3], K. Floettmann[4], F. Lemery[4]

[1]CANDLE SRI, 31 Acharyan Street, Yerevan 0040, Armenia
[2]Yerevan State University, 1 Alex Manoogyan Street, Yerevan 0025, Armenia
[3]A. Alikhanian National Scientific Laboratory (Yerevan Physics Institute)
2 Alikhanian Brothers Street, Yerevan 0036, Armenia
[4]Deutsches Elektronen-Synchrotron DESY, Notkestraße 85, 22607 Hamburg, Germany

**Abstract**- An algorithm for calculating the radiation field of a charged point particle performing a spiral motion in an infinite cylindrical waveguide with a multilayer side wall is found. The number of layers and their filling is arbitrary. The axis of the spiral is aligned with the axis of the waveguide, so that the geometry of the problem has cylindrical symmetry. Explicit expressions for modal frequency distributions and equations for resonant frequencies for single-layer and double-layer waveguides are given. Examples of graphical constructions of modal frequency distributions of modes for single-layer (resistive), double-layer (metal-dielectric) and triple-layer (metal-dielectric with internal NEG coating) waveguides are presented.

## 1. INTRODUCTION

Placing a helical undulator in a cylindrical waveguide with a metal wall transforms its radiation spectrum from continuous to discrete [1-5], which can expand its scope of application, allowing, by selecting waveguide and undulator parameters, to obtain narrow-band and narrow-beam radiation, and to optimize its radiation by establishing a single-mode regime [5].

The radiation spectrum of a particle moving along a spiral trajectory in a cylindrical waveguide with ideally conducting walls has the form of discretely located infinitely thin spectral lines and has singularities at critical points [4]. Replacing the ideal waveguide with a waveguide with resistive walls allows the elimination of the specified singularity [6]. The solution in [6] contains an indefinite function, the form of which is selected from the condition of coincidence of the limit transition of the obtained solution to the existing solution for an ideal waveguide [4]. In the present work, as a component of the solution for a waveguide with a multilayer wall, the recently obtained exact solution for the radiation of a particle moving along a spiral trajectory in free space [7] is used, which eliminates the need for artificial introduction of an additional function.

The use of a waveguide with a multilayer wall can improve the radiation properties: the addition of an internal dielectric layer weakens the mode attenuation, and an additional NEG gasket serves to maintain a high vacuum in the waveguide.

## 2. STATEMENT OF THE PROBLEM

The problem of determining the radiation field of a particle moving along a helical trajectory in an infinite cylindrical waveguide with a multilayer wall is considered. Initially, it is assumed that the axes of the boundary cylindrical surfaces of the layers that make up the waveguide wall coincide with the common axis of the waveguide.

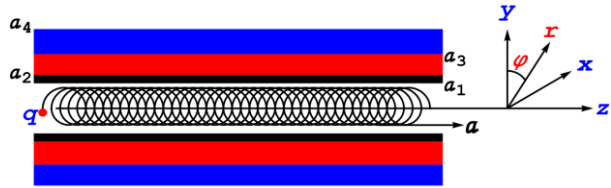

Figure 1: Multilayer cylindrical waveguide with a particle moving along a helical trajectory

The contents of each layer consist of isotropic metallic or dielectric materials. The number of layers N is arbitrary, but finite. The electromagnetic properties of the layer materials are described by their relative permittivity and magnetic permeability $\varepsilon_i, \mu_i$ $(i = 1,2,...N)$, with their arbitrary dependence on the frequency ω. The geometric shape of the structure is fixed by setting the internal radius of the waveguide $a_1$ and the radii of the cylindrical surfaces $a_i$ ($i = 2,3, ... N + 1$), separating the adjacent layers. The outer layer has an infinite thickness: it can be a medium with characteristics $\varepsilon_{N+1}, \mu_{N+1}$ or a vacuum with relative permittivity and magnetic permeability $\varepsilon_0 = \mu_0 = 1$. When considering the problem, a cylindrical coordinate system $r, \varphi, z$ is used, combined with the waveguide axis (Fig. 1).

The method of partial regions is used to solve the problem. In this case, each of the layers that make up the wall is taken as a separate region. Previously, a similar method was widely used to solve problems of radiation of linearly moving particles in cylindrical waveguides with single-layer [8] and multilayer [9] walls. An attempt to generalize the method to the case of particle motion along a helical trajectory was made in [6, 10]. The final version of the solution to this problem is given in this paper.

The standard technique for constructing a solution to the system of inhomogeneous Maxwell equations by the partial domain method in the case of a waveguide with layered walls consists of constructing partial solutions in each of the selected partial domains using elementary solutions of Maxwell's equations in a cylindrical coordinate system and stitching their tangential electric and magnetic components on cylindrical surfaces that separate adjacent domains (adjacent layers of the waveguide wall), including the internal vacuum cavity of the waveguide ($r \leq a_1$) and the external infinite layer ($r \geq a_{N+1}$) filled with vacuum, a dielectric, or a finitely conducting medium.

The algorithm for constructing radiation fields of a particle performing linear motion parallel to the axis of a cylindrical waveguide with a multilayer wall was first derived in [9]. We present its universalized modification, convenient for constructing radiation fields of particles with linear and spiral trajectories, determining their resonant frequencies, as well as the eigenvalues of the free oscillation fields in a multilayer waveguide.

The complete solution of the inhomogeneous Maxwell equations contains the sum of the general solution of the homogeneous Maxwell equations with undetermined weighting coefficients and a particular solution of the inhomogeneous Maxwell equations containing currents and charges generated by a moving particle.

## 3. GENERAL SOLUTION OF HOMOGENEOUS MAXWELL EQUATIONS

The components of the general solution are the electric $\vec{E}^{(i)}$ and magnetic $\vec{H}^{(i)}$ components of the frequency-time representations of partial waves propagating in each of the layers that make up the wall of the waveguide:

$$\vec{E}^{(i)} = \begin{cases} \sum_{m=-\infty}^{\infty} \vec{E}_m^{(i,J)}, & i = 0 \\ \sum_{m=-\infty}^{\infty} \{\vec{E}_m^{(i,J)} + \vec{E}_m^{(i,H)}\}, & i = 1, 2 ... N, \\ \sum_{m=-\infty}^{\infty} \vec{E}_m^{(i,H)}, & i = N + 1 \end{cases} \quad \vec{H}^{(i)} = \begin{cases} \sum_{m=-\infty}^{\infty} \vec{H}_m^{(i,J)}, & i = 0 \\ \sum_{m=-\infty}^{\infty} \{\vec{H}_m^{(i,J)} + \vec{H}_m^{(i,H)}\}, & i = 1, 2 ... N, \\ \sum_{m=-\infty}^{\infty} \vec{H}_m^{(i,H)}, & i = N + 1 \end{cases} \quad (1)$$

where $Z = J, H$ and

$$\vec{E}_m^{(i,Z)} = A_m^{(i,Z)} \vec{E}_{m,TM}^{(i,Z)} + B_m^{(i,Z)} \vec{E}_{m,TE}^{(i,Z)},$$
$$Z_0 \vec{H}_m^{(i,Z)} = A_m^{(i,Z)} \vec{H}_{m,TM}^{(i,Z)} + B_m^{(i,Z)} \vec{H}_{m,TE}^{(i,Z)}. \quad (2)$$

Here $A_m^{(i,Z)}$ and $B_m^{(i,Z)}$ are arbitrary weight coefficients before TM and TE components $\vec{E}_{m,TM}^{(i,Z)}$, $\vec{H}_{m,TM}^{(i,Z)}$ and of $\vec{E}_{m,TE}^{(i,Z)}$, $\vec{H}_{m,TE}^{(i,Z)}$ partial waves. If $Z = J$, $Z_m = J_m(v_{m,i} r)$, (Bessel function of the first kind) and if $Z = H$, $Z_m = H_m^{(1)}(v_{m,i} r)$ (Hankel function of the first kind); $Z_0 = 120\pi \, \Omega$ is the impedance of free space;

$$\vec{E}_{m,TM}^{(i,Z)} = -v_m^{(i)^{-2}} \operatorname{rot} \vec{R}, \quad Z_0 \vec{H}_{m,TM}^{(i,Z)} = j\varepsilon_i k v_m^{(i)^{-2}} \vec{R},$$
$$Z_0 \vec{H}_{m,TE}^{(i,Z)} = -v_m^{(i)^{-2}} \operatorname{rot} \vec{R}, \quad \vec{E}_{m,TE}^{(i,Z)} = -j\mu_i k v_m^{(i)^{-2}} \vec{R},$$
$$\vec{R} = \vec{e}_z \times \vec{\nabla} P^Z, \quad P^Z = Z_m e^{j(n\varphi + pz - \omega t)} \quad (3)$$

In (3) $p$ is the longitudinal eigenvalue of the $m^{th}$ mode. For a given mode it is the same for all layers; $v_m^{(i)}$ is the transverse eigenvalue of the $m^{th}$ mode in the $i^{th}$ layer. Its relationship with the transverse eigenvalue of the same mode in the internal vacuum cavity ($k = \omega/c$, $\omega$ is the frequency, $c$ is the speed of light in vacuum, $v_i = v_{m,i}$, $i = 0, 1, 2 \ldots N + 1$) is:

$$v_{m,i}^2 = k^2 \varepsilon_i \mu_i - p^2 = k^2(\varepsilon_i \mu_i - 1) + v_{m,0}^2 \tag{4}$$

Here $v_{m,0}$ is the transverse eigenvalue of the m-th harmonic in the vacuum cavity of the waveguide.

The homogeneous system of equations (in the absence of charges and currents in the cavity of the waveguide), obtained by matching the tangential field components at the boundaries, reduces to a system of four equations (similar to [10]):

$$\widehat{D}(k, p, v_{m,0}) \cdot \widehat{X} = 0, \tag{5}$$

$$\widehat{X} = \left\{ A_m^{(0,J)}, B_m^{(0,J)}, A_m^{(N+1,H)}, B_m^{(N+1,H)} \right\}$$

where $\widehat{X}$ is a four-element single-column matrix and $\widehat{D}(k, p, v_{m,0})$ is a four-column square matrix:

$$\widehat{D}(k, p, v_{m,0}) = \widehat{Q}\widehat{W}_H - \widehat{W}_J, \qquad \widehat{Q} = \prod_{i=1}^{N} \widehat{Q}_i, \tag{6}$$

The matrix $\widehat{W}_J$ corresponds to the values of the tangential components of the fields on the inner surface of the vacuum cavity of the waveguide (see (1) for $i = 0$ and $r = a_1$), and the matrix $\widehat{W}_H$ describes the tangential components of the fields on the boundary of the outer infinite layer (see (1) for $i = N + 1$ and $r = a_{N+1}$):

$$\widehat{W}_J = \begin{pmatrix} \alpha_{11} & 0 & 0 & 0 \\ \alpha_{21} & \alpha_{22} & 0 & 0 \\ 0 & \alpha_{11} & 0 & 0 \\ -\alpha_{22} & \alpha_{21} & 0 & 0 \end{pmatrix}, \quad \widehat{W}_H = \begin{pmatrix} 0 & 0 & \beta_{13} & 0 \\ 0 & 0 & \beta_{23} & \beta_{24} \\ 0 & 0 & 0 & \beta_{13} \\ 0 & 0 & \beta_{43} & \beta_{23} \end{pmatrix} \tag{7}$$

The harmonic fields in the vacuum cavity of the waveguide are characterized by the absence of components diverging on the waveguide axis (containing Hankel functions of the first kind), and when describing the fields in the outer infinite region, only components satisfying the radiation condition at infinity are preserved (there are no Bessel functions of the first kind). This explains the presence of columns with zero elements in (7). Explicit expressions for the nonzero elements of these matrices are given below:

$$\alpha_{11} = J_m(a_1 v_{m,0}), \quad \alpha_{21} = \frac{mp J_m(a_1 v_{m,0})}{a_1 v_{m,0}^2}, \quad \alpha_{22} = -j\frac{k J_m'(a_1 v_{m,0})}{v_{m,0}}$$

$$\beta_{13} = H_m^{(1)}(a_1 v_{m,N+1}), \quad \beta_{23} = -\frac{mp H_m^{(1)}(a_1 v_{m,N+1})}{a_{N+1} v_{m,N+1}^2}, \quad \beta_{24} = -j\mu_{N+1} \frac{k H_m^{(1)'}(a_1 v_{m,N+1})}{v_{m,N+1}}, \quad \beta_{43} = -\frac{\varepsilon_{N+1}}{\mu_{N+1}} \beta_{24} \tag{8}$$

The elements of the matrix $\widehat{Q}_i$ contain geometric ($a_i, a_{i+1}$) and electromagnetic ($\varepsilon_i, \mu_i$) parameters corresponding to the i-th layer with a finite thickness ($i = 1, 2, \ldots N$). It is represented using two two-dimensional matrices $\hat{\alpha}$ and $\hat{\beta}$ (9) and contains 6 independent elements (10):

$$\widehat{Q}_i = \frac{j\pi}{2} \begin{Bmatrix} \hat{\alpha} & \hat{\beta} \\ -\frac{\varepsilon_i}{\mu_i}\hat{\beta} & \hat{\alpha} \end{Bmatrix}, \quad \hat{\alpha} = \begin{Bmatrix} q_{11} & 0 \\ q_{21} & q_{22} \end{Bmatrix}, \quad \hat{\beta} = \begin{Bmatrix} q_{13} & q_{14} \\ q_{23} & -\frac{a_{i+1}}{a_i} q_{13} \end{Bmatrix} \tag{9}$$

$$q_{11} = a_{i+1} v_{m,i} U_{12}, \quad q_{13} = j\frac{mp}{\varepsilon_i k} U_{11}, \quad q_{14} = j\frac{a_{i+1} v_{m,i}^2}{\varepsilon_i k} U_{11}$$

$$q_{21} = \frac{mp}{a_i v_{m,i}} (a_i U_{21} - a_{i+1} U_{12}), \quad q_{22} = a_{i+1} v_{m,i} U_{21}, \quad q_{23} = j\frac{a_{i+1}}{k v_{m,i}^2} \left( k^2 \mu_i v_{m,i}^2 U_{22} - \frac{m^2 p^2}{a_i a_{i+1} \varepsilon_i} U_{11} \right) \tag{10}$$

The explicit form of the functions included in (10) is presented in (11). They have a form typical for similar problems (see [11, 9]):

$$U_{11} = J_m(a_{i+1} v_{m,i}) H_m^{(1)}(a_i v_{m,i}) - J_m(a_i v_{m,i}) H_m^{(1)}(a_{i+1} v_{m,i})$$

$$U_{12} = J_m'(a_{i+1} v_{m,i}) H_m^{(1)}(a_i v_{m,i}) - J_m(a_i v_{m,i}) H_m^{(1)'}(a_{i+1} v_{m,i})$$

$$U_{21} = J_m'(a_i v_{m,i}) H_m^{(1)}(a_{i+1} v_{m,i}) - J_m(a_i v_{m,i+1}) H_m^{(1)'}(a_i v_{m,i})$$

$$U_{22} = J_m'(a_i v_{m,i}) H_m^{(1)'}(a_{i+1} v_{m,i}) - J_m'(a_i v_{m,i+1}) H_m^{(1)'}(a_i v_{m,i}) \tag{11}$$

The following identity holds:

$$\widetilde{U}_i = U_{12}U_{21} - U_{11}U_{22} = -4/(\pi^2 v_{m,i}^2 a_i a_{i+1}) \tag{12}$$

When $a_i = a_{i+1}$, matrix (9) degenerates into a unit diagonal matrix: in this case $U_{11} = U_{22} = 0$, and $U_{12} = U_{21} = -2j/\pi a_i v_{m,i}$. For $p = \sqrt{k^2 - v_{m,0}^2}$ the equation

$$\widehat{D}\left(k, \sqrt{k^2 - v_{m,0}^2}, v_{m,0}\right) = 0 \tag{13}$$

defines the complex transverse eigenvalue $v_{m,0}$ as a function of k, i.e., is a dispersion relation. Examples of constructing dispersion curves for single-layer (resistive), two-layer (metal-dielectric) and three-layer (metal-dielectric-NEG) can be found in [10] and [12].

In the presence of a particle moving in a waveguide along a rectilinear trajectory parallel to the axis with a velocity v, we should set $p = \omega/v$, $v_{m,0} = \sqrt{k^2 - p^2} = \omega\sqrt{1/c^2 - 1/v^2} = j\omega/v\gamma$ ($\gamma$ is the Lorentz factor of the particle), then the equation

$$\widehat{D}\left(\frac{\omega}{c}, \frac{\omega}{v}, j\frac{\omega}{v\gamma}\right) = 0 \tag{14}$$

will determine the complex resonance frequencies of the wake radiation during rectilinear motion of the particle. If the particle moves along a helical trajectory (longitudinal component of velocity $v_z$, rotation frequency $\omega_0$), then we should set [7] $p = (\omega - m\omega_0)/v_z$ and $v_{m,0} = \sqrt{(\omega/c)^2 - (\omega - m\omega_0)^2/v_z^2}$. In this case, the equation defining the complex resonance frequencies is:

$$\widehat{D}\left(\frac{\omega}{c}, (\omega - m\omega_0)/v_z, \sqrt{(\omega/c)^2 - (\omega - m\omega_0)^2/v_z^2}\right) = 0 \tag{15}$$

Note that the resonance frequencies do not depend on the displacement of the trajectory from the axis in the case of a linear trajectory of the particle and on the radius of the spiral in the case of helical motion of the particle.

## 4. PARTICULAR SOLUTION OF INHOMOGENEOUS MAXWELL EQUATIONS

As a particular solution of the inhomogeneous Maxwell equations, the solution for the radiation field of a particle moving along an infinite helicoidal trajectory in free space, obtained in [7], is taken. The particular solution, as a general one, is obtained by the method of partial regions. In this case, the space is divided into two regions: the region outside the cylinder containing the helicoidal trajectory of the particle ($r > a$) and the region inside this surface ($r < a$):

$$\vec{\mathcal{E}} = \begin{cases} \sum_{m=-\infty}^{\infty} \vec{\mathcal{E}}_m^{(J)} \\ \sum_{m=-\infty}^{\infty} \vec{\mathcal{E}}_m^{(H)} \end{cases}, \quad \vec{\mathcal{H}} = \begin{cases} \sum_{m=-\infty}^{\infty} \vec{\mathcal{H}}_m^{(J)} & r < a \\ \sum_{m=-\infty}^{\infty} \vec{\mathcal{H}}_m^{(H)} & r > a \end{cases}. \tag{16}$$

Each term of the multipole expansion is represented as a superposition of TM $\left(\vec{E}_{m,TM}^{(Z)}\right)$ and TE $\left(\vec{E}_{m,TE}^{(Z)}\right)$ modes, with $Z = J$ or $H$, and arbitrary weight factors $\left(\mathcal{A}_m^{(Z)}, \mathcal{B}_m^{(Z)}\right)$, i. e. it is composed of fundamental solutions of the homogeneous Maxwell equations in cylindrical coordinates:

$$\begin{aligned} \vec{\mathcal{E}}_m^{(Z)} &= \mathcal{A}_m^{(Z)} \vec{E}_{m,TM}^{(Z)} + \mathcal{B}_m^{(Z)} \vec{E}_{m,TE}^{(Z)} \\ Z_0 \vec{\mathcal{H}}_m^{(Z)} &= \mathcal{A}_m^{(Z)} \vec{H}_{m,TM}^{(Z)} + \mathcal{B}_m^{(Z)} \vec{H}_{m,TE}^{(Z)} \end{aligned}, \tag{17}$$

where

$$\begin{aligned} \vec{E}_{m,TM}^{(Z)} &= -v_m^{-2} \operatorname{rot} \vec{R}, \quad Z_0 \vec{H}_{m,TM}^{(Z)} = jkv_m^{-2} \vec{R}, \\ Z_0 \vec{H}_{m,TE}^{(Z)} &= -v_m^{-2} \operatorname{rot} \vec{R}, \quad \vec{E}_{m,TE}^{(Z)} = -jkv_m^{-2} \vec{R} \, . \\ \vec{R} &= \vec{e}_z \times \vec{\nabla}\left(Z_m e^{j(m\varphi + pz - \omega t)}\right). \end{aligned} \tag{18}$$

As before, if $Z = J$, i.e., $r < a$, $Z_m = J_m(v_m r)$ and if $Z = H$, i.e. $r > a$, $Z_m = H_m^{(1)}(v_m r)$, where $J_m$ and $H_m^{(1)}$ are the Bessel function and the Hankel function of the first kind. In (18) $p_m$ and $v_m = \sqrt{\omega^2/c^2 - p_m^2}$ are the longitudinal and transverse eigenvalues of the $m^{\text{th}}$ mode. In the case of a linear motion of the particle $p_m = \omega/v$ (v is total velocity of the particle) and $v_m = j\omega/v\gamma$ (j is the imaginary unit), while for the helical motion:

$$p_m = (\omega - m\omega_0)/v_z, \quad v_m = v_{m,0} = \sqrt{\omega^2/c^2 - (\omega - m\omega_0)^2/v_z^2}. \tag{19}$$

where $v_z$ is longitudinal component of the particle velocity and $\omega_0$ is a rotational frequency of the particle.

The amplitudes $\mathcal{A}_m^{(Z)}$ and $\mathcal{B}_m^{(Z)}$ are determined using the boundary conditions [13], which determine the discontinuity of the fields on the surface r=a, containing charges and currents caused by the motion of the particle along this surface [7]:

$$\begin{aligned}\mathcal{A}_m^{(J)} \\ \mathcal{A}_m^{(H)}\end{aligned} = Z_0 \frac{qc}{4v_z\omega} \left(\frac{cv_m}{\pi\omega}\right)^{1/2} f_m \begin{cases} H_m^{(1)}(av_m), \\ J_m(av_m) \end{cases}, \quad \begin{aligned}\mathcal{B}_m^{(J)} \\ \mathcal{B}_m^{(H)}\end{aligned} = jZ_0 \frac{aq\omega_0}{4} \left(\frac{cv_m}{\pi\omega}\right)^{1/2} v_m \begin{cases} H_m^{(1)\prime}(av_m), & r < a \\ J_m'(av_m), & r > a \end{cases}$$

$$f_m = \omega(\omega_0 m - \omega/\gamma_z^2), \quad \gamma_z^2 = (1 - v_z^2/c^2)^{-1} \tag{20}$$

## 5. COMPLETE SOLUTION OF INHOMOGENEOUS MAXWELL'S EQUATIONS

Taking into account the particular solution (16)-(20) transforms equation (5) into a linear inhomogeneous equation (with non-zero right-hand sides):

$$\widehat{D}(k, p_m, v_{m,0}) \cdot \widehat{X} = \widehat{S} \tag{21}$$

The arguments $p_m, v_{m,0}$ in (21) correspond to the notations (12); $\widehat{S}$ is a single-column four-element matrix, the elements of which are the values of the tangential electric and magnetic components of the fields (17) on the inner surface of the waveguide $r = a_1$:

$$\widehat{S} = \left\{\mathcal{E}_{m,z}^{(H)}, \mathcal{E}_{m,\varphi}^{(H)}, Z_0\mathcal{H}_{m,z}^{(H)}, Z_0\mathcal{H}_{m,\varphi}^{(H)}\right\}_{/r=a_1} \tag{22}$$

The solution of equation (21), which determines the sought amplitudes $A_m^{(0,J)}, B_m^{(0,J)}, A_m^{(N+1,H)}, B_m^{(N+1,H)}$, is

$$\widehat{X} = \widehat{D}^{-1}(k, p_m, v_{m,0}) \cdot \widehat{S} \tag{23}$$

where $\widehat{D}^{-1}(k, p_m, v_{m,0})$ is the matrix inverse to the matrix $\widehat{D}(k, p_m, v_{m,0})$. Thus, the complete solution of the inhomogeneous Maxwell equations, describing the radiation of a particle moving along a helicoidal trajectory in a cylindrical waveguide with a multilayer wall, in the vacuum cavity of the waveguide ($r < a$) is:

$$\vec{E}_m^{(0)} = A_m^{(0,J)} \vec{E}_{m,TM}^{(0,J)} + B_m^{(0,J)} \vec{E}_{m,TE}^{(0,J)} + \begin{cases} \mathcal{A}_m^{(J)} \vec{E}_{m,TM}^{(J)} + \mathcal{B}_m^{(J)} \vec{E}_{m,TE}^{(J)}, & r < a \\ \mathcal{A}_m^{(H)} \vec{E}_{m,TM}^{(H)} + \mathcal{B}_m^{(H)} \vec{E}_{m,TE}^{(H)}, & r > a \end{cases}$$

$$Z_0 \vec{H}_m^{(0)} = A_m^{(0,J)} \vec{H}_{m,TM}^{(0,J)} + B_m^{(0,J)} \vec{H}_{m,TE}^{(0,J)} + \begin{cases} \mathcal{A}_m^{(J)} \vec{H}_{m,TM}^{(J)} + \mathcal{B}_m^{(J)} \vec{H}_{m,TE}^{(J)}, & r < a \\ \mathcal{A}_m^{(H)} \vec{H}_{m,TM}^{(H)} + \mathcal{B}_m^{(H)} \vec{H}_{m,TE}^{(H)}, & r > a \end{cases} \tag{24}$$

The radiation field in the outer infinite layer ($r > a_{N+1}$) is written as follows:

$$\vec{E}_m^{(N+1)} = A_m^{(N+1,H)} \vec{E}_{m,TM}^{(N+1,H)} + B_m^{(N+1,H)} \vec{E}_{m,TE}^{(N+1,H)}$$

$$Z_0 \vec{H}_m^{(N+1)} = A_m^{(N+1,H)} \vec{H}_{m,TM}^{(N+1,H)} + B_m^{N+1,H} \vec{H}_{m,TE}^{(N+1,H)} \tag{25}$$

If the outer layer is filled with vacuum ($\varepsilon_{N+1} = \mu_{N+1} = 1$), then field (25) can be interpreted as radiation emanating from the outer wall of the waveguide into the surrounding space.

## 6. SPECIAL CASES

In this section, explicit expressions are given for the amplitudes $A_m^{(0,J)}, B_m^{(0,J)}$ of the radiation fields of a particle in a vacuum cavity of a waveguide for two-layer ($N = 2$) and single-layer ($N = 1$) walls. They are conveniently expressed through the amplitudes $\mathcal{A}_m^{(H)}, \mathcal{B}_m^{(H)}$ (20) of a particular solution:

$$A_m^{(0,J)} = \frac{A_1 \mathcal{A}_m^{(H)} + B_1 \mathcal{B}_m^{(H)}}{A_3}, \quad B_m^{(0,J)} = \frac{B_2 \mathcal{A}_m^{(H)} + A_2 \mathcal{B}_m^{(H)}}{A_3}, \tag{26}$$

The inner layer of the wall of a two-layer waveguide with electromagnetic characteristics $\varepsilon_1, \mu_1$ has a thickness $d_1 = a_2 - a_1$,, while the outer layer with characteristics $\varepsilon_2, \mu_2$ is infinite: $a_2 \leq d_2 \leq \infty$. The coefficients $A_i$ and $B_i$ included in (26) are represented in the form of polynomials in powers of $m$:

$$A_i = A_i^{(0)} + A_i^{(2)} m^2 + A_i^{(4)} m^4, \quad i = 1,2,3$$
$$B_i = B_i^{(1)} m + B_i^{(3)} m^3, \quad i = 1,2 \tag{27}$$

The explicit form of the expansion coefficients is given below:

$$\begin{Bmatrix} A_1^{(0)} \\ A_2^{(0)} \\ -A_3^{(0)} \end{Bmatrix} = -a_1^2 a_2^2 k^4 v_0^2 v_1^4 v_2^2 \begin{Bmatrix} S_\varepsilon^H S_\mu^J \\ S_\varepsilon^J S_\mu^H \\ S_\varepsilon^J S_\mu^J \end{Bmatrix}, \quad \begin{Bmatrix} A_1^{(2)} \\ A_2^{(2)} \\ -A_3^{(2)} \end{Bmatrix} = k^2 p^2 v_1^2 \begin{pmatrix} X \\ X \\ X_1 \end{pmatrix} + a_1^2 v_0^2 H_2^2 v_{12}^2 \begin{Bmatrix} L_{\varepsilon,1}^H L_{\mu,1}^J \\ L_{\mu,1}^H L_{\varepsilon,1}^J \\ L_{\mu,1}^J L_{\varepsilon,1}^J \end{Bmatrix}$$

$$A_1^{(4)} = A_2^{(4)} = -\frac{H_1}{J_1} A_3^{(4)} = -H_1 J_1 H_2^2 p^4 U_{11}^2 v_{01}^2 v_{12}^2$$

$$B_1^{(1)} = -B_2^{(1)} = \frac{2}{\pi} a_1 k^3 p v_0 v_1^4 v_2 (s + u)$$
$$s = a_1 \varepsilon_1 \mu_1 H_2^2 \tilde{U}_1 v_0^2 v_{12}, \quad u = a_2 v_{01} M_\varepsilon^H M_\mu^H \tag{28}$$

$$B_1^{(3)} = -B_2^{(3)} = -\frac{2}{\pi a_2 v_2} a_1 H_2^2 k p^3 U_{11}^2 v_0 v_1^2 v_{01} v_{12}^2$$

In (28) the following notations are introduced:

$$\begin{Bmatrix} S_{\varepsilon(\mu)}^H \\ S_{\varepsilon(\mu)}^J \end{Bmatrix} = v_1 \varepsilon_2(\mu_2) H_2' \begin{Bmatrix} L_{\varepsilon(\mu),1}^H \\ L_{\varepsilon(\mu),1}^J \end{Bmatrix} - v_2 \varepsilon_1(\mu_1) H_2 \begin{Bmatrix} L_{\varepsilon(\mu),2}^H \\ L_{\varepsilon(\mu),2}^J \end{Bmatrix}$$

$$\begin{Bmatrix} L_{\varepsilon(\mu),j}^H \\ L_{\varepsilon(\mu),j}^J \end{Bmatrix} = v_0 U_{2j} \varepsilon_1(\mu_1) \begin{Bmatrix} H_1 \\ J_1 \end{Bmatrix} + v_1 U_{1j} \begin{Bmatrix} H_1' \\ J_1' \end{Bmatrix}, j = 1,2$$

$$\begin{Bmatrix} X \\ X_1 \end{Bmatrix} = a_2 v_2^2 v_{01} \{2s + u\} \begin{Bmatrix} H_1 \\ J_1 \end{Bmatrix} J_1, \quad \begin{Bmatrix} M_\varepsilon^H \\ M_\mu^H \end{Bmatrix} = v_1 U_{11} H_2' \begin{Bmatrix} \varepsilon_2 \\ \mu_2 \end{Bmatrix} - v_2 U_{12} H_2 \begin{Bmatrix} \varepsilon_1 \\ \mu_1 \end{Bmatrix}$$

$$H_2 = H_m^{(1)}(a_2 v_2), \quad H_1 = H_m^{(1)}(a_1 v_0), \quad J_1 = J_m(a_1 v_0)$$

$$v_i = v_{m,i}, \quad v_{ij} = v_i^2 - v_j^2 \tag{29}$$

To go from a two-layer waveguide to a single-layer one, it is sufficient to set $a_1 = a_2$ in Equations (28), (29). In this case, the Equations are significantly simplified:

$$\begin{Bmatrix} A_1 \\ A_2 \end{Bmatrix} = -a_2 k^2 v_0^2 v_2^2 \begin{Bmatrix} K_\varepsilon^H K_\mu^J \\ K_\varepsilon^J K_\mu^H \end{Bmatrix} + H_2^2 J_1 H_1 m^2 p^2 v_{02}^2$$

$$A_3 = a_2 k^2 v_0^2 v_2^2 K_\varepsilon^J K_\mu^J - H_2^2 J_1^2 m^2 p^2 v_{02}^2, \quad B_1 = -B_2 = \frac{1}{\pi} H_2^2 m p v_2^2 v_{02}^2 \tag{30}$$

where

$$\begin{Bmatrix} K_{\varepsilon(\mu)}^H \\ K_{\varepsilon(\mu)}^J \end{Bmatrix} = H_2' \varepsilon_2(\mu_2) v_0 \begin{Bmatrix} H_1 \\ J_1 \end{Bmatrix} - H_2 v_2 \begin{Bmatrix} H_1' \\ J_1' \end{Bmatrix}. \tag{31}$$

As a result, we have a waveguide with an internal radius $a_2$ and with an infinite wall, the material of which is characterized by electric and magnetic permeability $\varepsilon_2, \mu_2$. If $\varepsilon_2 = 1 + j Z_0 \sigma/k$ ($\sigma$ is the conductivity of the wall material) and $\mu_2 = 1$, we have a waveguide with an infinite resistive wall.

## 7. NUMERICAL EXAMPLES

As is known [1-6], the frequency distributions of the radiation fields of a particle in a waveguide with ideally conducting walls have a discrete character. The radiation fields in a waveguide with a metal wall, the material of which has a high but finite conductivity (a waveguide with a resistive wall), have the same properties. In an ideal waveguide, the distribution consists of infinitely thin spectral lines fixed at a discrete set of resonant frequencies. In a resistive waveguide, as well as in waveguides with layered walls, the spectral lines have a finite width and a fixed amplitude at the resonant frequency. Both in free space [7] and in an ideal waveguide, and in waveguides with layered walls (with an arbitrary number of layers), the allowed frequency band of the m-th term of the multipole expansion is determined by the inequality

$$\frac{m\omega_0}{1+v_z^2/c^2} \leq \omega \leq \frac{m\omega_0}{1-v_z^2/c^2} \tag{32}$$

($v_z$ is the longitudinal component of the particle velocity), determined from the condition $Re\{v_{0,m}\} > 0, Im\{v_{0,m}\} = 0$. The geometry of the spiral trajectory of the particle is determined by three parameters: the total v and longitudinal $v_z$ velocities of the particle (assumed to be constant) and the spiral period $l$. In this case, the orbital velocity of the particle $v_\varphi$, the rotation frequency $\omega_0$ and the orbital radius $a$ are determined by the formulas $v_\varphi = \sqrt{v^2 - v_z^2}$, $\omega_0 = 2\pi v_z/l$, $a = v_\varphi/\omega_0$. In the numerical examples considered below, $v = 0.99c, v_z = 0.98c, l = 5cm$ and $a_1 = 1cm$. The conductivity of the outer infinite wall is taken to be equal to $\sigma = 58 \cdot 10^6 \Omega^{-1} m^{-1}$ (copper).

Figures 2-6 (a) show the distributions of the amplitudes $A_m^{(0,J)}$, $B_m^{(0,J)}$: $m = 1,2$ for a single-layer resistive waveguide (Fig. 2, 3) and $m = 1$ for waveguides with two- and three-layer walls (Fig. 3-6). For comparison, the distributions of the amplitudes $\mathcal{A}_m^{(H)}, \mathcal{B}_m^{(H)}$ of the radiation field of a spirally moving particle in free space are given, which we use as a particular solution of the inhomogeneous Maxwell equations and which is an integral part of the complete solution. The same Figures 2-6 (b) show the corresponding frequency distributions of the radial component of the field for a certain value of the radial coordinate ($r = 0.5a_1$) with and without taking into account the particular solution. The amplitude distributions contain resonances corresponding to certain modes of the cylindrical waveguide with the corresponding number of layers in the wall. The amplitude distributions, calculated using the above-developed method, correspond to TM and TE modes separately. The hybrid nature of the modes is manifested in the presence of certain bursts in TM modes at resonant frequencies corresponding to TE modes, and vice versa. When constructing the field components (Figures 2-6, (b)), the amplitudes and the corresponding synchronous (located at the same frequencies) bursts form a superposition with weighting coefficients determined by the expressions for the fields (3). The resonant frequencies of the local maxima of the distributions remain unchanged.

Let us note one feature characteristic of all five presented examples: the modules of the TM and TE amplitudes have equal-sized bursts (Fig. 2-6, a) at the extreme points of the allowed region (32). They are marked with crosses on the graphs. As can be seen from the same Figures located on the right, they are mutually compensated when constructing the field components. For a particle performing a helicoidal motion in free space, the mutual conjugacy of the TM and TE modes is proven analytically. Here we limit ourselves to a graphical demonstration.

In the case of a single-layer resistive waveguide (Fig. 2, 3), the high conductivity of copper leads to values of the dimensionless transverse eigenvalues $v_{0,1}a_1$ (Fig. 2) and $v_{0,2}a_1$ (Fig. 3) that are close, at resonant frequencies, to the transverse eigenvalues of a waveguide with perfectly conducting walls, i.e., to the roots of the Bessel functions (for TM modes) or to the roots of its derivative (for TE modes). The transverse eigenvalues are determined by substituting the values of the resonant frequencies into formula (19). For the first term of the multipole expansion ($m = 1$), the resonant values of the dimensionless transverse eigenvalues are close to the roots of the first-order Bessel functions or their derivatives (Fig. 2), and for the second term of the multipole expansion ($m = 2$), they are close to the roots of the second-order Bessel functions or their derivatives. The numbering (or qualification) of the hybrid modes of a resistive waveguide is carried out accordingly: the conventional designation ($TM_{mn}$) ($TE_{mn}$) corresponds to modes whose normalized transverse eigenvalues are close to the n-th root of the Bessel function of the $m$-th order (or its derivative). From Figures 2 and 3 it follows that there is a pairwise coincidence: each mode with a certain transverse eigenvalue excited in the low-frequency region ($f < f_0$) corresponds to a mode with a close transverse eigenvalue located in the high-frequency region ($f > f_0$). This is a consequence of the Doppler effect: the part of its radiation directed forward acquires a high frequency, while the low-frequency part of the radiation is directed backward. In an ideal waveguide, both branches (low-frequency and high-frequency) of the same mode have equal transverse eigenvalues [4]. The presence of a resistive wall (Fig. 2, 3) and additional dielectric (Fig. 4) and metallic (Fig. 5, 6) layers leads to differences in their values. The high-frequency branch of the spectral distribution, corresponding to the radiation directed forward, is most sensitive to distortions caused by the resistivity of the wall and additional layers. In the case of linear motion of a particle in a resistive waveguide, the spectral distribution of its radiation (impedance) is a smooth curve with a broadband resonance [14]. If the wall of the resistive waveguide is coated on the inside with a thin dielectric layer, the smooth curve is transformed into a curve with a single narrowband resonance (or multiple narrowband resonances, if the dielectric coating is thick enough) [15]. In the case of helical motion of a particle, its radiation in a resistive waveguide already has a resonant character. Adding a thin dielectric layer does not change the number of resonances and does not significantly affect their location and amplitude characteristics (Fig. 3): there is an insignificant shift in the resonances of the lower-order modes corresponding to forward radiation ($f > f_0$), towards low frequencies, which leads to an increase in their transverse eigenvalues. The positive effect of the additional dielectric layer is manifested in the possibility of a significant reduction in the exponential attenuation decrement of forward-directed radiation, which can be achieved by optimally selecting the permittivity and thickness of the dielectric layer (see [10]). The attenuation decrement is found by solving equation (15), which determines a discrete set of complex resonant frequencies for a given term of the multipole expansion, the real parts of which correspond to the real resonant frequencies, and their imaginary components are proportional to the values of the attenuation decrement.

In Figures 2-6 (a), in addition to the distributions of the amplitudes $A_m^{(0,J)}$, $B_m^{(0,J)}$, which serve as weighting coefficients for the field components caused by the general solution, the distributions of the amplitudes $\mathcal{A}_m^{(H)}, \mathcal{B}_m^{(H)}$, included in the particular solution of the inhomogeneous Maxwell's equations are also shown. Accordingly, the Figures 2-6 (b) show the frequency distributions of the radial electric components of the radiation field caused by both the full and the general solutions. The contribution of the particular solution to the general picture of the amplitude and field distributions can be either insignificant (Figs. 2-4) or significant (Figs. 5,6). In the first case, the elements of the particular solution introduce a weak homogeneous background into the general picture, which does not have a significant effect on it. In the second case, the background is on the same level with the amplitudes and field distributions of the general solution, which distorting and suppressing the forward-directed radiation.

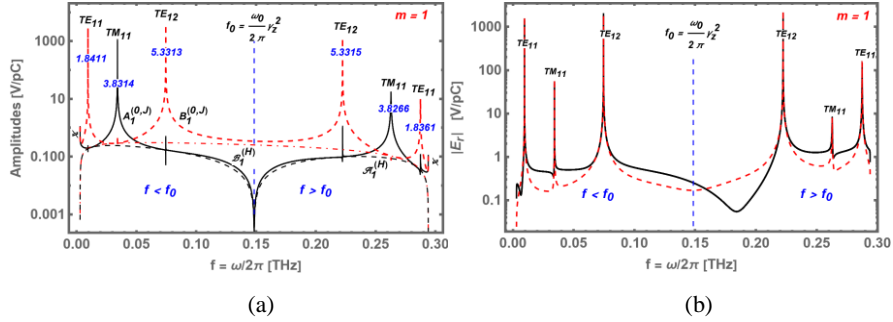

Figure 2: Spectral distribution of amplitudes (a) and radial electric component at $r = 0.5 a_1$ (b) in a single-layer resistive waveguide; $m = 1$.

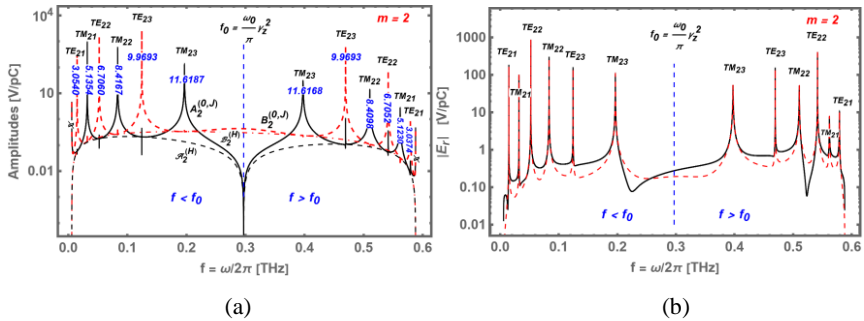

Figure 3: Spectral distribution of amplitudes (a) and radial electric component at $r = 0.5 a_1$ (b) in a single-layer resistive waveguide; $m = 2$.

An additional NEG layer applied over the dielectric (Fig. 5, 6) is designed to eliminate residual gas molecules in the waveguide and maintain a high vacuum in it. Two types of NEG materials were used in the calculations: with conductivities $\sigma_1 = 1.4 \cdot 10^4 \Omega^{-1} m^{-1}$ (Fig. 5) and $\sigma_2 = 8 \cdot 10^5 \Omega^{-1} m^{-1}$ (Fig. 6), developed at PSI and intended for coating the vacuum chambers of the SLS-2 storage ring [16]. Their effect on the properties of radiation of a particle performing linear motion in a three-layer waveguide was studied in [12]. It was found that at $\sigma = \sigma_1$ (note that $\sigma_1 < \sigma_2$) the NEG coating does not have a destructive effect on the resonance properties of the radiation even at a fairly large thickness ($\sim 10\mu m$), whereas at $\sigma = \sigma_2$ the resonance nature of the radiation is disrupted even at very small thicknesses of the NEG layer ($\sim 0.1\mu m$). It is obvious that other regularities take place during the helical motion of the particle, which are yet to be established. Perhaps a thinner layer of the NEG coating should be used.

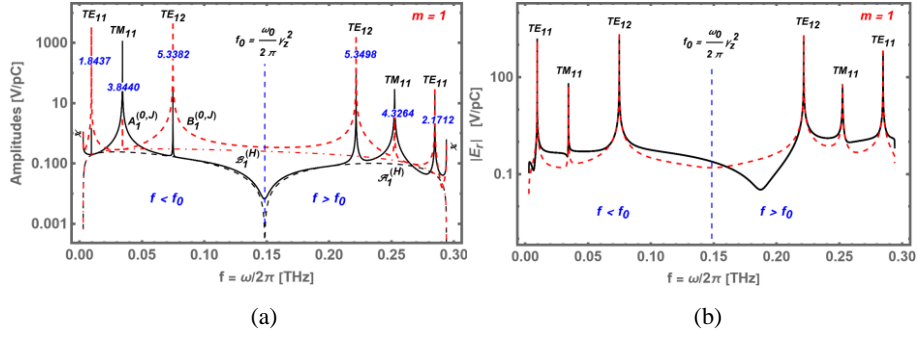

Figure 4: Spectral distribution of amplitudes (a) and radial electric component at $r = 0.5a_1$ (b) in a two-layer copper-dielectric waveguide; thickness of the dielectric layer $d = a_2 - a_1 = 10\mu m$; permittivity $\varepsilon_1 = 10$, $\mu_1 = 1$; $m = 1$.

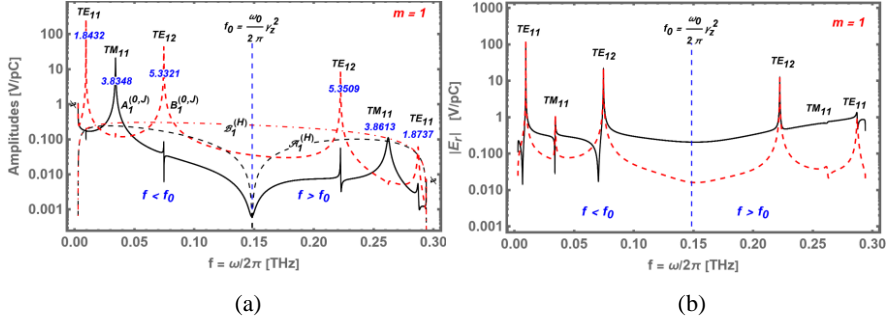

Figure 5: Spectral distribution of amplitudes (a) and radial electric component at $r = 0.5a_1$ (b) in a three-layer copper-dielectric waveguide with internal NEG coating; thickness of the dielectric layer $d_2 = a_3 - a_2 = 10\mu m$; permittivity $\varepsilon_2 = 10\mu m$, $\mu_2 = 1$; thickness of NEG coating $d_1 = a_2 - a_1 = 1\mu m$; conductivity $1.4 \cdot 10^4 \Omega^{-1} m^{-1}$; $m = 1$.

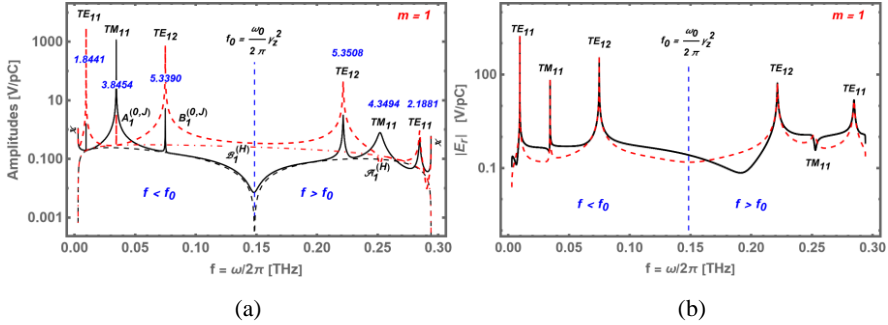

Figure 6: Spectral distribution of amplitudes (a) and radial electric component at $r = 0.5a_1$ (b) in a three-layer copper-dielectric waveguide with internal NEG coating; thickness of the dielectric layer $d_2 = a_3 - a_2 = 10\mu m$; permittivity $\varepsilon_2 = 10\mu m$, $\mu_2 = 1$; thickness of NEG coating $d_1 = a_2 - a_1 = 1\mu m$; conductivity $8 \cdot 10^5 \Omega^{-1} m^{-1}$; $m = 1$.

## 8. CONCLUSION

The main content of this paper is the presentation of an algorithm for determining the radiation field of a particle moving along a spiral trajectory in a cylindrical waveguide with a multilayer wall. An important component of this algorithm is the exact solution for the radiation of a particle moving along a spiral trajectory in free space, first obtained in [7]. Numerical examples demonstrate its effectiveness in constructing fields and reveal some of their features, in particular, the narrow-resonance nature of the fields in a single-layer resistive and two-layer metal-dielectric waveguide, a certain pattern is visible in the formation of the values of resonant frequencies and transverse eigenvalues when passing from a resistive to a metal-dielectric waveguide. Naturally, it is impossible to conduct an exhaustive analysis based on the five examples given. It is necessary to conduct an in-depth analysis to determine the optimal geometric and electromagnetic parameters of the layers of two- and three-layer waveguides according to several criteria: maximum power and narrow-band radiation at resonant frequencies, minimization of the distorting effect of the NEG coating, minimization of attenuation decrements, the possibility of establishing a single-mode radiation regime, etc. Some results remained outside the scope of this article: for example, it can be shown that the phase velocities of both branches of the mode field generated by a particle with a spiral trajectory are synchronous with the longitudinal component of the particle velocity $v_z$.

## ACKNOWLEDGMENTS

The work was supported by the Science Committee of RA in the frame of the research project 23SC-CNR-1C006